\documentclass[prc,aps,tightenlines,floatfix,showpacs,twocolumn]{revtex4}
\usepackage[dvips]{graphics,color}
\usepackage{graphicx}
\usepackage{dcolumn}
\usepackage{bm}
\usepackage{epsfig}
\usepackage{supertabular}
\usepackage{amsmath}

\newcommand{\bd}[1]{ \mbox{\boldmath $#1$}}

\newcommand{\beq}{\begin{equation}}
\newcommand{\eeq}{\end{equation}}
\newcommand{\beqa}{\begin{eqnarray}}
\newcommand{\eeqa}{\end{eqnarray}}

\begin{document}
\def\ii{\'\i}

\title{
Renormalization of coherent state variables, within the geometrical
mapping of algebraic models
}

\author{ H. Y\'epez-Mart\ii nez$^1$,
G. Morales-Hern\'andez$^2$,
P. O. Hess$^{2,3,4}$, G. L\'evai$^5$ and P.R. Fraser$^{6,7}$  \\
{\small\it
$^1$Universidad Aut\'onoma de la Ciudad de M\'exico,
Prolongaci\'on San Isidro 151,} \\
{\small\it
Col. San Lorenzo Tezonco, Del. Iztapalapa,
09790 M\'exico D.F., Mexico} \\
{\small\it
$^2$ Instituto de Ciencias Nucleares, UNAM, Circuito Exterior, C.U.,} \\
{\small\it A.P. 70-543, 04510 M\'exico, D.F., Mexico} \\
{\small\it $^3$Frankfurt Institute for Advanced Studies, Johann Wolfgang Goethe Universit\"at,} \\
{\small\it Ruth-Moufang-Str. 1, 60438 Frankfurt am Main, Germany} \\
{\small\it $^4$GSI Helmholtzzentrum f\"uer Schwerionenforschung GmbH,}\\
{\small\it Max-Planck-Str. 1, 64291 Darmstadt, Germany}\\
{\small\it
$^5$Institute of Nuclear Research of the
Hungarian Academy of Sciences,} \\
{\small\it
Debrecen, Pf. 51, Hungary-4001} \\
{\small\it
$^6$ Instituto Nazionale di Fisica Nucleare, Sezione di Padova,
I-35131, Italy \\
$^7$ School of Physics, University of Melbourne, Victoria 3010, Australia
} \\
}

\begin{abstract}
We investigate the geometrical mapping of algebraic models. As
particular examples we 
 consider 
the
{\it Semimicriscopic Algebraic Cluster Model} (SACM) and the
{\it Phenomenological Algebraic Cluster Model} (PACM), 
 which also contains
the vibron model
 as a special case. 
In the
geometrical mapping coherent states are employed as trial states.
We show that the coherent state variables have to be renormalized
and not the interaction terms of the Hamiltonian, as is usually done.
The coherent state variables will depend on the total number of
bosons and the coherent state variables.
The nature of these
variables is extracted through a relation obtained by
comparing physical observables,
such as the distance between the clusters or the quadrupole deformation
of the nucleus, to their algebraic counterpart.
\end{abstract}

\maketitle

\section{Introduction}
\label{one}

Algebraic models play an important role in nuclear and particle
physics in 
 understanding  
collective properties of nuclei
\cite{iba}, clusters of nuclei \cite{vibron,sacm1,sacm2,who,barrett},
atomic molecules \cite{levine} and the spectrum of hadrons
\cite{bijker}, just to name a few examples.
In order to achieve a more 
 physically inspired interpretation 
it is useful to employ a geometrical mapping,
using coherent states \cite{roosmalen1,roosmalen2,kirson}.
More recent investigations on geometrical mappings can be found
in 
 Refs.~
\cite{cejnar,jolie,casten-2009,casten-2010}.
The semi-classical potential is determined by the expectation value
of the Hamiltonian with respect to the normalized coherent state.
 To the best of our knowledge Ref. \cite{scholten} was the first 
 work in which coherent states were applied within an algebraic 
 model, namely the {\it Interacting Boson Approximation} (IBA). 

In 
 establishing a relation with  
geometric quantities the following procedure
is usually applied:
Denoting generically the coherent state variables by $\alpha$, 
the expectation value of the one-, two- and three-body
interactions with respect to the coherent state 
 is obtained as 
$N\alpha^2$, $N(N-1)\alpha^4$ and
$N(N-1)(N-2)\alpha^6$, respectively 
\cite{roosmalen1,roosmalen2,cejnar,jolie,scholten}.
The $\alpha$ variable
is then interpreted to be proportional (in lowest order) to a
physical variable, for example the deformation $\beta$ of a nucleus
\cite{iba} or some other coordinate \cite{roosmalen1,jolie,cejnar}.
The dependence 
 on 
$N$ is used as an argument to renormalize the
different order of interactions \cite{roosmalen1,roosmalen2,cejnar,jolie}.
 Apparently 
the first 
 application of  
this procedure 
 to 
atomic and nuclear
 systems 
was in
 Refs.~
\cite{roosmalen1,roosmalen2}, where the expectation values, with respect
to the coherent state, of the one- and two-body interactions in the
vibron model were divided by $N$ and $N(N-1)$ respectively,
with the argument that "it is convenient" to define an intensive ${\bd H}$.
To us, this argument is not conclusive nor justified,
 on grounds 
that intensive
quantities are defined by dividing an operator globally by $N$, 
 rather than dividing 
individual terms 
in an expansion 
independently 
by 
 different powers of 
$N$.
The expectation values,
divided
usually 
by appropriate powers in $N$,
were set equal to $\alpha^2$ and $\alpha^4$, which in turn were assumed
to be of the order of one and identified with physical coordinates, such as
the relative distance between two mass points. 
 (The
momentum dependence
was 
 also 
discussed, using complex coherent state variables. 
For simplicity, in order to describe the problem and the solution to it, we are only
interested in the coordinate dependence and will set the imaginary part
of the cartesian components of $\alpha$ equal to zero.)
Later on
 a general rule became accepted according to which 
one-, two- and
three-body interactions are divided by $N$, $N(N-1)$ and $N(N-1)(N-2)$,
respectively. This dependence was also suggested in
\cite{scholten} in the early years of the IBA.

 Reference \cite{gilmore-book} gives a 
clearer insight into why the former procedure is applied. 
We suspect that
the origin of the
renormalization of the interaction parameters is found in
chapter 15 of \cite{gilmore-book}, where systems of
$N$ {\it identical} bosons or fermions are discussed, and thus
we provide here a summary of material therein. 

The Hamiltonian is a
function of the generators of a Lie algebra (see also \cite{gilmore-1979}), 
 i.e. 
${\bd H}({\bd T}_i)$, with ${\bd T}_i$ as 
 short-hand  
notation for the
generators.
The thermodynamical limit of
a Lie algebra is obtained by dividing each generator by $N$
(i.e., one defines ${\bd T}_i/N$) giving a new
generator. The newly defined generators commute in the limit
$N \rightarrow \infty$ 
 and 
thus represent classical operators.
This limiting procedure corresponds to a {\it contraction} of a Lie
algebra and is well described in 
 Ref.~
\cite{gilmore-lie}.
A model
Hamiltonian is then given as a function in ${\bd T}_i/N$
and additionally multiplied by $N$. The Hamiltonian has the structure
${\bd H}=N{\bd h}({\bd T}_i/N)$, and thus 
a scaling of the Hamiltonian
in $N$ is {\it additionally} imposed.
As a special example, the Meshkov-Glick-Lipkin (MGL) and the Dicke model
are discussed. Factors in powers of $1/N$ are introduced in order to maintain
the contribution of a higher order interaction on the same level as 
 that of 
a one-body
interaction. The scaling of the interaction terms
is necessary, because 
in a two-level model the lowest level is completely occupied and when the
number of states in the lowest level is increased, so 
 is 
$N$.
 Without 
this renormalization of the interaction parameters 
the two-body interaction would completely dominate the
one-body interaction, resulting
 a useless model. 
In other words, the proposed scaling recipe is
introduced by necessity. 

A further argument
 Ref. \cite{gilmore-lie} provides 
is that
in nuclear physics a clear scaling of the binding energy
is observed and it should be reproduced in realistic models. As logical as
this arguments sounds, in models of nuclear physics, like the shell model,
one does not always have a scaling in the interaction terms as a function
in the number of nucleons. For example,
in order to obtain the scaling of the binding energy, the {\it Liquid Drop
Model} \cite{eisenbergI} is applied. 
 To incorporate 
corrections due to the shell
structure,
the Strutinsky method 
 is used 
(see, for example,
\cite{eisenberg}). The
average behavior of the calculated energy, as a function on the number of
nucleons, is subtracted and the result, containing the fluctuations
due to the shell structure, is then added to the
Liquid Drop Model. Thus, in the shell model itself no scaling of the
Hamiltonian appears.

Another example is the geometric model of the nucleus
\cite{eisenbergI,GCM}. The basic degrees of freedom are spin two bosons,
related to the surface vibration and rotation of the nucleus.
As in the IBA \cite{iba} one can introduce a spin zero boson and impose
the condition that the total number of bosons $N$ is fixed. The
cutoff is removed by letting $N$ go towards large values. In this
geometric model no rescaling of the interaction terms in inverse powers in
$N$ is applied, because this would annihilate the higher n-body interaction
compared to the one-body interaction, thus eliminating important structures.

This shows that a model Hamiltonian  does not necessarily
require a renormalization of the interaction parameters. As we 
 shall 
show
further below, the interaction parameters of the Hamiltonians,
discussed in this contribution, will depend
on a cutoff and approach 
fixed values for a large cutoff. It shows that
the rules 
given in 
 Refs.~
\cite{gilmore-book,gilmore-1979} are not always
 valid 
and depend on the model itself. Applying 
these rules 
 indiscriminately 
may lead
to inconsistent results, though the rules are justified with regards to certain
 types  
of models.
As we will see, inconsistencies become particular noticeable for models
where the $N$ has to approach infinity (removing a cut-off).
In section II we first summarize the definitive features of all algebraic
Hamiltonian methods, as well as some problems in the conventionally-used
renormalization procedure used for such.
In section III we discuss the geometric mapping.  
In section IV we propose an alternative 
 procedure that 
is based on the relation of algebraic 
 operators 
to their geometric 
 counterparts. 
This will 
 relate 
the coherent state parameters $\alpha$ to a physical 
 quantity. 
These variables will depend on $N$ and are approximately proportional to
the 
 given
physical 
 quantity. 
The $N$-dependent factors in front of 
 an 
n-body
interaction will disappear in the large $N$ 
 limit, so 
no renormalization
of the $n$-body interaction 
 will be required.  
 Instead 
a renormalization of the coherent
state variables, $\alpha$,
 becomes necessary. 
We have the impression that the reason for the confusion is related to
 treating 
the 
 $N$-dependent 
factors and the coherent state variables ($\alpha$) 
as of different origin
and therefore independent. However, they are not.
As the main examples of an algebraic model, we 
 shall consider 
the 
 Semimicroscopic Algebraic Cluster Model (SACM) and the 
 Phenomenologic Algebraic Cluster Model (PACM). 
After 
 discussing 
the results in section V,
conclusions are drawn 
 in section VI.

\section{The structure of algebraic Hamiltonians}
\label{two}

An algebraic Hamiltonian of boson systems, which includes, 
in general, 
different types of bosons, exhibits the following dependence
\beqa
{\bd H} & = & {\bd H} ( {\bd a}^\dagger, {\bd a}, {\bd b}^\dagger, {\bd b},
..., p)
~~~,
\label{eq1}
\eeqa
where the ${\bd a}^\dagger_{i_a}$, ${\bd a}_{i_a}$,
${\bd b}^\dagger_{i_b}$, ${\bd b}_{i_b}$, etc.,
are boson creation and annihilation operators,
satisfying the usual commutation relations. The indices $i_a$, $i_b$, ...
vary from 1 to $d_a$, $d_b$, ...,
 denoting 
the degrees of freedom
for each type of boson. We associate 
a definite
angular momentum $l_a$, $l_b$, etc. 
 to each boson. 
{\it Thus
the total number of degrees of freedom is}
$d_a + d_b + ...=d$, where 
$d_k=(2l_k+1)$ ($k=a,b,...$).
The $p$ in (\ref{eq1}) is a 
 short-hand 
notation
for the interaction parameters $(p_k)$. The Hamiltonian is 
 constructed as 
an $SO(3)$ scalar,
i.e., all operators are coupled to angular momentum zero.
These are examples where there are different 
 kinds 
of particles, 
 as opposed to those in Refs. 
\cite{gilmore-book,gilmore-1979}.

The interactions are usually divided 
 into 
one-, two-, three- body and higher
 terms. 
The one-body interactions, coupled to angular momentum
zero, are given by
\beqa
{\bd n}_a & = & \sum_{i_a} {\bd a}^\dagger_{i_a} {\bd a}_{i_a} ~,~
{\bd n}_b ~ = ~ \sum_{i_b} {\bd b}^\dagger_{i_b} {\bd b}_{i_b} ~,~ ...~~~.
\label{eq2}
\eeqa
These are the respective number operators. In algebraic boson models,
one requires that the {\it total number of bosons is constant}, i.e.,
\beqa
N & = & {\bd n}_a + {\bd n}_b + ... ~=~ {\rm const}
~~~.
\label{eq3}
\eeqa
A general one-body interaction is then given by
\beqa
\sum_k \epsilon_k {\bd n}_k
~~~.
\label{eq3.1}
\eeqa
Similar considerations can be given for 
 higher-order 
interactions.

In 
 Ref.~
\cite{jolie}, which gives a 
 comprehensive 
resum\'e on the until-now
accepted procedure, the different kinds of bosons are denoted by the
same letter ${\bd b}$, but with indices 
 that distinguish 
between different
 kinds 
of bosons. At least one boson has angular momentum zero, which we also
 shall 
adopt 
 and 
identify the $a$-boson with this scalar boson.
This implies that $i_a$=1,  
 so this 
index 
can be skipped,
for convenience.

The notation 
 set up 
above 
is generic, 
and 
 it contains all the 
special cases. 
For example, the IBA \cite{iba}, which
describes 
quadrupole excitations of the nucleus in its basic
version, has two types of bosons, namely a scalar $s$-boson and a quadrupole
$d$-boson. The total number of bosons, $N=n_s+n_d$, is given by half
the number of nucleons in the valence shell and is therefore constant
and {\it finite}.

The notation is also valid for
the 
 Semimicroscopic Algebraic Cluster Model (SACM) 
\cite{sacm1,sacm2} 
 and the Phenomenologic Algebraic Cluster Model (PACM), 
 which contains 
the {\it Vibron Model} 
 as a special case, i.e. that in which the clusters are closed-shell 
 nuclei  
\cite{paper-I,paper-II}. 
The 
 former 
observes the Pauli exclusion principle 
 between all the nucleons, including those belonging to different clusters, 
while the
latter does not. In 
 Refs. 
\cite{paper-I,paper-II} this difference was
investigated and 
the geometric mapping was 
 also 
performed 
 in order to 
study 
quantum phase transitions.
The basic degrees of freedom in the SACM and PACM are the relative
oscillation quanta, described by spin one $\pi$-bosons. These bosons
are the shell model oscillation quanta. In order that the algebraic
Hamiltonian conserves the total number of bosons, a scalar $\sigma$-boson
is introduced, which has no physical meaning;
 rather, its role is merely to 
introduce a cutoff. This cutoff has to tend to infinity, which, in general,
requires 
 $n_{\pi} << n_{\sigma}$. 
In some dynamical symmetry limits
($SO(4)$ in the PACM), the numbers
may be of the same order. The PACM is also applied to molecules, where
the Pauli exclusion principle is not required.

The main point here is
that there are different kinds of bosons, which carry different angular
momentum. They have different number of degrees of freedom, depending on
the model and system. The number of bosons of a certain type,
contained in a given eigenstate of the Hamiltonian, may be
quite different 
 from the number of other type bosons. 
Coming back to the one-body
interactions of Eq. (\ref{eq2}), they are proportional to the
respective 
 boson numbers,  
i.e.,
\beqa
{\bd n}_a & \rightarrow & n_a ~,~ {\bd n}_b ~ \rightarrow ~ n_b ~,~...
\label{eq4}
\eeqa
and {\it not} to $N$ as suggested, for example, in \cite{jolie}.

In the $U(5)$ dynamical symmetry limit of the IBA, the number of 
$s$- and $d$-bosons
is well defined but, in general quite different. The number operator of
 these bosons 
is proportional to 
 $n_s$ and $n_d$, respectively. 
 The corresponding number operators of the SACM and PACM are 
proportional to 
 $n_\sigma$ and $n_\pi$,
respectively. Furthermore, because
the total number of bosons has to be large and in general, 
$n_\pi <<N$,
i.e. $n_\sigma >> n_\pi$, the matrix elements of the corresponding
number operators are very different 
 from those 
suggested in 
 Ref. 
\cite{jolie}.

For the two-body interactions (this is the last example discussed and
it should be clear how 
 higher-order 
interactions behave), the matrix
elements of interactions, which depend only on the $a$-bosons 
 and $b$-bosons, are proportional to 
 $n_a(n_a-1)$ 
 and 
$n_b(n_b-1)$, 
etc.,  
 respectively. 
It is not 
 difficult 
to imagine that two-body
interactions, which depend on two types of boson, behave as
$n_an_b$, etc. Again,
using coherent states
they do not scale as $N(N-1)$, as suggested in
\cite{cejnar}, but rather as $N(N-1)\alpha^4$, i.e., the $N$ and
$\alpha$ dependence are linked together and 
in the scaling consideration one has to treat both at the same time.

The question is: 
How can the inconsistencies be resolved?
In order to propose a possibility, we have to discuss coherent states 
used
for geometrical mappings. After that, we 
 shall
analyze the origin of
 the inconsistencies
of the old renormalization procedure and 
 discuss 
how 
 it should be modified to make it consistent. 

\section{The geometrical mapping using coherent states}

The general structure of a coherent state, used as a trial state
in an algebraic model, is as follows:
\beqa
\mid \alpha , {\bd \beta}, ... \rangle & = &
{\cal N} \left( \alpha a^\dagger + (\beta \cdot b^\dagger )
+...\right)^N
\mid 0 \rangle
 ~~~,
\label{eq5}
\eeqa
 where 
${\cal N}$ is an easy to determine normalization factor,  
 and the vacuum is denoted by $\mid 0 \rangle$. 
We assumed
that the $a$-boson is the scalar one and the dot indicates
a scalar product. The 
$\alpha$ can be factorized out, as done in some
models. 

Examples of coherent states are: 
\beqa
\mid {\bd \beta} \rangle & = & \frac{1}{\sqrt{N!(1+\beta^2)}}
\left[ s^\dagger + ({\bd \beta} \cdot d^\dagger )\right]^N \mid 0 \rangle
\label{eq6}
\eeqa
 for the IBA and  
\beqa
\left| \alpha \right> &=& 
{\cal N}_{N,n_0}
(\boldsymbol{\alpha} \cdot \boldsymbol{\pi}^\dagger)^{n_0}
\left[
\boldsymbol{\sigma}^\dagger 
+ \left( \boldsymbol{\alpha} \cdot \boldsymbol{\pi}^\dagger
\right)\right]^{N} \left| 0 \right>
\nonumber \\
&=& {\cal N}_{N,n_0} \frac{N!}{(N+n_0)!}
\frac{{\rm d}^{n_0}}{{\rm d}\gamma^{n_0}_1}
\left[
\boldsymbol{\sigma}^\dagger 
+ \gamma_1 \left(\boldsymbol{\alpha} \cdot \boldsymbol{\pi}^\dagger
\right)\right]^{N+n_0} \left| 0 \right> 
\nonumber \\ 
\label{eq7}
\eeqa
 for the SACM, 
where 
 following Refs. \cite{paper-I} 
we redefined the total number of relative
oscillation quanta as 
\linebreak
$(N+n_0)$, while the $\gamma_1$ parameter has to
be set equal to 1 after the differentiation. 
 Here $n_0$ denotes the minimal number of $\pi$ bosons as required 
 by the Pauli principle. In the PACM $n_0=0$ and the 
coherent state reduces to 
 that of 
the
more commonly known vibron model \cite{frank}.

In the IBA a relation between the $\beta$ deformation variable and the one 
of the
geometric model \cite{eisenberg} is given in Ref.~\cite{ginocchio-NPA1980}.
This 
 relation is based on  
comparing the quadrupole density 
 obtained in the two models. 
As a result, the physical
quadrupole deformation, ${\bar \beta}$, turns out to be roughly proportional to
$\left( \frac{2N}{A} \right)\beta$, where $N$ is the total number of bosons
and $A$ is the mass number of a nucleus. These values are much lower than
$\beta$. This demonstrates that the naive interpretation of $\beta$ being
identical to the value of the quadrupole deformation is incorrect and the
correct one can only be found by comparing the expectation value
of one operator in the geometric model to the equivalent one in the
algebraic model.
Additionally, as already mentioned, the identification of the geometric variable
$\beta_{{\rm phys}}$
has been obtained for a given $N$ together with the coherent state variable $\beta$,
and not independently.

For further illustration, let us 
 consider 
 the expectation value of the $\pi$-boson number operator
in the PACM, 
 i.e. with $n_0=0$. 
(This 
 quantity 
does not depend
on an interaction parameter.) The result is \cite{geom}
\beqa
\langle \boldsymbol{n}_\pi \rangle & = & N \frac{\alpha^2}
{\left( 1+\alpha^2\right)} ~~~.
\label{eq8}
\eeqa
We know that this expectation value is equal to the average number
of $\pi$-bosons, which in the $SU(3)$ dynamical symmetry limit is
$n_\pi$. Thus, it 
 cannot 
be of the order of $N$, but rather
\beqa
N\frac{\alpha^2}{(1+\alpha^2)} & = & n_\pi
~~~,
\label{eq9}
\eeqa
which is of the order $n_\pi << N$. This suggests that the coherent state
variable $\alpha$ has to be renormalized, as already
indicated in the last section.

As just indicated, in order to obtain a consistent relation of the
coherent state variables to physical geometrical variables, one has
to rely on the comparison of the expectation value of 
 one 
(or more)
{\it algebraic} operator(s), with respect to a trial state (here a
coherent state) to the corresponding {\it geometric} expression of
the operator(s). This comparison gives a relation of the coherent
state parameter 
 with 
the corresponding geometrical variable. 

\section{Renormalization of the coherent state variable $\alpha$}

 Here we give a more detailed discussion of the ideas outlined above 
 for the SACM and the PACM.  
 The corresponding treatment of the IBA can be found in Ref.  
\cite{ginocchio-NPA1980}, 
 and the ideas can be transferred to other algebraic models. 

The basic conditions we require for the Hamiltonian and the
coherent state variables are that
for large $N$ \\
i) the Hamiltonian 
 cannot 
depend on 
an arbitrary 
cut-off, represented by the total
number of bosons $N$, at least in the limit for very large $N$, and \\
ii) the parameter $\alpha$ of the coherent state has to be related to an
independent operator, as for example the distance operator.

 Condition i) introduces the following requirement for the $N$-dependence 
 of the  
interaction parameters in the Hamiltonian. 
Let 
 $p(n)$ symbolize 
the parameter of 
 an 
n-body
interaction. In the SACM this parameter, once it is adjusted to
experiment, should acquire
a fixed value $p_0(n)$ for $N \rightarrow \infty$. When the cutoff $N$ is
chosen to be finite, the interaction parameter has 
a different value 
 in general. The dependence on $N$ can be parametrized in
powers of $1/N$, i.e.,
\beqa
p(n) & = & p_0(n) + \frac{p_1(n)}{N} + \frac{p_2(n)}{N^2}+...
~~~
\label{eq10}
\eeqa
Applying an
adjustment to experimental data, with increasing $N$, the parameter $p(n)$
approaches $p_0(n)$. No renormalization of the interaction is
required. In the IBA, the "renormalization", given by
$p(2)/\left[ N(N-1)\right]$ is equivalent to 
 writing 
 the parameter
$p^\prime (n)$ instead, because $N$ is finite and fixed. This implies that it
is a matter of convenience to include the $N$-dependent factor or not.
This is quite different in the cluster models, where $N$ is
{\it arbitrary} but fixed. When the cutoff is removed, $N$ has to tend
to infinity. This difference is the main reason we insist on the example
of the cluster models.

With respect to ii), we first define an algebraic
distance operator, which for
$N \rightarrow \infty$ approaches the usual distance operator
(this discussion can also be found in Ref.~\cite{geom}):
\beqa
\boldsymbol{r}_m & = & \sqrt{ \frac{\hbar}{2m\omega}}
\left( \boldsymbol{\pi}^\dagger_m + \boldsymbol{\pi}_m \right)
~~~.
\label{eq11}
\eeqa
When the $\sigma$-bosons are introduced, this operator has to be
changed. The new operator, called the {\it algebraic coordinate
operator}, should satisfy the following minimal conditions: \\
a) The
total number of bosons has to be kept constant, i.e., each
$\boldsymbol{\pi}_m^\dagger$ has to be multiplied by
$\boldsymbol{\sigma}$ and each $\boldsymbol{\pi}_m$ has to be
multiplied by a $\boldsymbol{\sigma}^\dagger$; \\
b) the definition of
the distance operator should be {\it independent} of the basis and
Hamiltonian used (though, it is permissible to give the representation
of the operator in one particular basis); and \\
c) for $N\rightarrow\infty$ it should converge
to the standard form given in (\ref{eq11}).  

The proposed algebraic
distance operator is given by
\beqa
\boldsymbol{r}_m^a  = \sqrt{ \frac{\hbar}{2Nm\omega}}
\left( \boldsymbol{\pi}^\dagger_m \boldsymbol{\sigma} +
\boldsymbol{\sigma}^\dagger \boldsymbol{\pi}_m \right)
~~~,
\label{eq12}
\eeqa
where ``$a$" refers to {\it algebraic}. The operator
itself leaves the total number of bosons unchanged, as required by 
condition a). The $N$ in the denominator of the square root is
introduced because in the harmonic oscillator basis the matrix
elements of the $\sigma$-operators are proportional to $\sqrt{N-n_\pi}$, which
for large $N$ and small number of $\boldsymbol{\pi}$ bosons is
approximated by $\sqrt{N}$. {\it This approximate value of the
$\boldsymbol{\sigma}$ operators is satisfied in any basis, with the
condition that the average number of $\boldsymbol{\pi}$ bosons is
much smaller than $N$} (though the structure is particularly simple
in the harmonic oscillator basis).  Thus the $1/\sqrt{N}$ factor
cancels approximately the contributions due to the addition of the
$\boldsymbol{\sigma}^\dagger$ and $\boldsymbol{\sigma}$ operators. In
this form, the algebraic distance operator does not depend on the
basis used ($\boldsymbol{r}^a_m$ can be applied to any kind of
basis), nor on the Hamiltonian or any particular dynamical symmetry,
thus satisfying condition b). For
very large $N$ the expressions of the physical and the algebraic
coordinate operators tend to each other, satisfying condition c).

This definition agrees with \cite{lemus1,lemus2}, where an algebraic
model for {\it atomic molecules} is discussed. Often (see for example
\cite{levit}) one defines the radial distance in terms of the
dynamical symmetry, relating it indirectly to the matrix element of
the dipole operator, without any further considerations. This violates
condition b) above. We 
 stress 
that the definition of the radial
distance operator has to be independent of the Hamiltonian in the
Hilbert space considered. The Hamiltonian determines 
 whether 
there is a
dynamical symmetry or not, which should be independent of the radial
distance operator, while the Hilbert space as such is independent of
the basis used.

The expectation value of the algebraic coordinate operator
with respect to the coherent state is
\beqa
\langle \boldsymbol{r}_m^a \rangle & = & \sqrt{ \frac{2N\hbar}{m\omega}}
\frac{\alpha_m}{\left(1+\alpha^2\right)} ~=~ r_m^a
~~~.
\label{eq13}
\eeqa
(Note that the $\alpha$ defined here is not the same as the $\alpha$
used in (\ref{eq5}).)

We define this as the algebraic distance $r_m^a$, which is, by
definition, of the order of one.  Inverting this relation gives

\beqa
\frac{\alpha_m}{\left(1+\alpha^2\right)} & = &
\sqrt{ \frac{m\omega}{2N\hbar}} r_m^a
~~~,
\label{eq14}
\eeqa
which again provides the dependence of $\alpha_m$ on $N$. 
This result suggests
to redefine $\alpha_m$ in terms of $\delta_m$ and $N$ as
\beqa
\alpha_m & = & \frac{\delta_m}{\sqrt{N}}
\label{eq15}
\eeqa
which gives

\begin{eqnarray}
\frac{\delta_m}{( 1+\frac{\delta^2}{N})} & = &
\sqrt{ \frac{m\omega}{2\hbar}} r_m^a
\label{eq16}
~~~,
\end{eqnarray}
a dimensionless measure of the distance between the two nuclear
clusters. For large $N$, the $\delta_m$ is 
 directly 
proportional to
$r_m^a$.
We claim that this is a consistent way to define the
separation between clusters trough the variable $\delta_m$.

The validity of (\ref{eq13}), the definition of $r_m^a$, depends on the
fluctuations of the related expectation value. 
The 
 intercluster 
distance vector can always be chosen along
the 
 z axis. 
The square of the variation ($\langle \boldsymbol{r}_0^{a
  2} \rangle - \langle \boldsymbol{r}_0^a \rangle^2$) can then also be
calculated, giving
\begin{eqnarray}
\langle (\boldsymbol{r}_0^{a})^{2} \rangle 
- \langle \boldsymbol{r}_0^a \rangle^2 & = &
-\left( \frac{4\hbar}{m\omega}\right) 
\frac{\left(\frac{\delta_0^2}{N}\right)}{\left(1+\frac{\delta^2}{N} \right)^2}
+ \left( \frac{\hbar}{2m\omega}\right)
\frac{1+{\textstyle\frac{\delta_0}{N}}}{1+{\textstyle\frac{\delta^2}{N}}}
\nonumber \\
& \rightarrow & \left( \frac{\hbar}{2m\omega}\right)
~~~,
\label{eq17}
\end{eqnarray}
also implying, in the second line, the large $N$ limit. 
As long as the
expectation value of the algebraic distance operator is greater than
the square root of this expression, it is safe to identify the $r_0^a$
as the distance between the two clusters. The square root of
(\ref{eq17}) gives numbers of the order of 1 fm. If $r_m^a$ is of
the order of the variation, it represents rather a {\it mean distance} of the
two clusters.

The above considerations have been, up to now, relevant
to the PACM, were no Pauli exclusion
principle is taken into account. In the SACM some further differences appear.
Because $n_\pi$ has a lower bound, known as the
Wildermuth condition \cite{wildermuth}, it was shown in \cite{geom}
that for $\alpha = 0$ there is
already a minimal distance between the clusters,
given by $r_0 \sim \sqrt{n_0}$. This can be understood easily by noting
that the maximal contribution of the radial wave function for non-zero
oscillation quanta is located at $r>0$.  
 Equation (\ref{eq14}) is the same, with the exception
that $r_m^a$ on the right hand side has to be substituted by
$\left( r_m^a -r_{0,m}\right)$, with $r_0^2 = \sum_m (-1)^m r_{0,m} r_{0,-m}$ 
 being 
the square of the minimal distance $r_0$ of the two clusters \cite{geom}.
Thus, the generally declared spherical limit, $\alpha = 0$, in the SACM
corresponds to a minimal distance between the two clusters.

\section{Discussion}

The question is now where the generally accepted procedure and our new
proposal produce important differences.
 Focusing 
only at the potential is
not always sufficient, because in the former
procedure the $n$-body interactions
are divided by powers in $N$, while in our new procedure these factors appear
in the redefinition of the coherent state variable, e.g. $\alpha$ in the
SACM and PACM. The 
 final 
results look the same and the discussion on
phase transitions give also the same results. Problems, however, appear
when $N \rightarrow \infty$.
When in the cluster model
the one-body interaction is divided by $N$, e.g., $\hbar\omega{\bd n}_\pi$
$\rightarrow$ $\frac{\hbar\omega}{N}{\bd n}_\pi$
(note that $\hbar\omega$ is the distance in energy between two shells and
is, therefore, fixed), for large $N$ the
excited states approach 
zero energy. If one does not divide
the one-body 
 term 
by $N$, one has at least to divide the two-body interaction
by $(N-1)$. Then, the argument repeats, i.e., for a pure $\pi$-boson
two-body interaction the matrix element scale as $n_\pi^2/(N-1)$, which
for $N \rightarrow \infty$ and $n_\pi << N$ approaches zero again.

With respect to the interaction parameters, the situation is
different for a finite $N$, as in the IBA. Adjusting the parameters to
experiment, one can increase the parameters of a two-body interaction such
that they compensate for the $(N-1)$ in the denominator.

Therefore, the
problem becomes only obvious in models like the SACM or PACM, were the
total number of bosons represent only a cut-off, which has to approach
infinity.

Another problem is related to the geometric values of observables, 
such as
the quadrupole deformation (or the distance between two clusters).
The principal argument is that the expectation value of an algebraic
observable would be put equal to the corresponding geometrical
observable, without 
 using 
any additional factors. For example,
if one determines the expectation
value of the quadrupole operator $Q_0$
(or the algebraic distance operator $r_m^a$)
with respect to a coherent state, its
value scales approximately like $N\alpha^2$. If one associates
to $\alpha$ a deformation, or distance, depending on the algebraic model,
the value of $Q_0$ ($r_m^a$) would increase 
 proportionally 
to $N$, i.e., it
would be too large. In \cite{roosmalen1,roosmalen2} these
expectation values are divided, without a clear justification,
by $N$ in order to get rid of the wrong $N$-dependence.
In other words, the geometrical interpretation runs into severe
problems, except when arbitrarily some $N$-dependent factors are
introduced.

\section{Conclusions}

In this contribution we discussed the geometrical mapping
of algebraic models, using coherent states, and the interpretation
of the coherent state variables. We showed that a consistent
procedure involves the renormalization of the coherent state variables and
not the many-body interaction parameters of the Hamiltonian.

This revised renormalization procedure introduces
{\it in the geometrical mapping} the factor
of $1/N$ for the 1-body interaction, $1/(N(N-1))$ for the 2-body interaction,
etc.
In conclusion, the renormalization procedure, as
proposed in \cite{jolie}, introduces the $N$-dependent factors in the
Hamiltonian, while the parameter $\alpha$ of the coherent state is taken
as of the order of one, resulting in a semi-classical potential which does
not depend on $N$ for large values of $N$. In contrast, we leave the
Hamiltonian untouched
and the 
 $N$-dependent 
factors enter
via the renormalization procedure for $\alpha$ as presented here.

As already mentioned in the introduction, the structure of the model
Hamiltonian depends on the system in consideration. In a two level model,
with all states in the lower level occupied, one has to include 
 $N$-dependent 
factors in order to get 
 reasonable 
results. In other models, like
the shell model and the ones presented here, the rules given in
\cite{cejnar,gilmore-book} lead to inconsistencies and another
renormalization procedure has to be applied. We also showed that 
 for 
$N=const.$ the renormalization of the interaction parameters does not affect
the results, 
 provided that 
the parameters are free and adjusted to a specific system.

Not following the revised procedure leads to a wrong and inconsistent
geometric interpretation and, at least, to conceptual difficulties.

\section*{Acknowledgements}

We gratefully acknowledge financial help from
DGAPA-PAPIIT
(no. IN103212),
from the National
Research Council of Mexico (CONACyT), OTKA (grant No. K72357), and
from the MTA-CONACyT joint project. POH acknowledges very useful
discussions with Octavio Casta\~nos (ICN-UNAM), related to the
differences in phase transitions in finite systems to the use of
coherent states. Also useful discussions with Roelof Bijker (ICN-UNAM)
are acknowledged, related to the definition of the radial distance.
The authors are also thankful to J\'ozsef Cseh for illuminating 
discussions on the subject.

\end{document}